\documentclass[11pt,oneside]{article}
\usepackage{a4wide}
\usepackage{mathrsfs}
\usepackage{latexsym,bm}
\usepackage{graphicx}
\usepackage{indentfirst}
\usepackage{slashed}
\usepackage{amsmath}
\usepackage{amssymb}
\usepackage{color}
\usepackage{hyperref}
\usepackage{epsfig}
\usepackage[titletoc]{appendix}
\usepackage{multirow}%
\usepackage{rotating}
\usepackage{cite}
\usepackage{ulem} 
\usepackage{extpfeil}
\usepackage[top=2.9cm,bottom=2.5cm,left=2.8cm,right=3cm]{geometry}
\usepackage{tabularx}
\usepackage{booktabs}
\usepackage{threeparttable}

\newcommand{\email}[1]{\footnote{{\em } \texttt{#1}}}

\newcommand{\rxt}{{\rm R}\chi{\rm T}}

\begin{document}
\thispagestyle{empty}
\title{
\Large \bf Revisit of tensor-meson nonet in resonance chiral theory } 
\author{\small Cheng~Chen,\, Nai-Qian Cheng,\, Lin-Wan Yan,\, Chun-Gui Duan,\, Zhi-Hui Guo\email{zhguo@hebtu.edu.cn} \\[0.3em]
{\small\it Department of Physics and Hebei Key Laboratory of Photophysics Research and Application, } \\
{\small\it Hebei Normal University,  Shijiazhuang 050024, China}
}

\date{}

%

\maketitle
\begin{abstract} 
We study the properties of the lowest multiplet of light-flavor tensor meson resonances, i.e. $f_2(1270)$, $a_2(1320)$, $K_2^*(1430)$, and $f_2'(1525)$, within the resonance chiral theory approach. The higher-order resonance chiral operators, including the light-quark mass and $1/N_C$ corrections,  are simultaneously incorporated in our study. The use of resonance chiral expressions allows us to analyze not only the relevant experimental data but also in the meantime the lattice results at unphysical quark masses, including the masses of the lowest multiplet of tensor resonances and their decay widths into two pseudoscalar mesons. In addition, the radiative decays of the tensor resonances into one photon plus one pseudoscalar meson and two photons are also  studied. 
\end{abstract}

\section{Introduction}

The lowest multiplet of light-flavor tensor meson resonances with $J^{P}=2^{+}$, consisting of $f_2(1270)$, $a_2(1320)$, $K_2^*(1430)$, and $f_2'(1525)$ that will be simply denoted as $f_2$, $a_2$, $K_2^*$ and $f_2'$ in order, are important objects in hadron physics. They play prominent roles in numerous physical processes, such as the $J/\psi \to \gamma P P'$ (being $P$ the light-flavor pseudoscalar mesons) decays~\cite{BESIII:2015rug,BESIII:2016gkg,BESIII:2018ubj}, the $\gamma^{(*)}\gamma^{(*)}\to P P'$ processes (which are in turn also important to pin down the theoretical uncertainties arising from the light-by-light scattering to  anomalous magnetic moment of muon)~\cite{Dai:2014zta,Danilkin:2016hnh,Gerardin:2017ryf}, the various photoproduction reactions off the nucleon target~\cite{Shklyar:2004ba,Yu:2011fv,Yu:2011zu}, the identification of the tensor glueball candidates~\cite{Bugg:1996by,Li:2000zb,Ye:2012gu,Li:2011xr}, and the yet unsettled issue about the tensor resonance contributions to the chiral low energy constants~\cite{Toublan:1995bk,Ecker:2007us}, etc.  

Unlike the opaque nature and somewhat controversial assignment of the light-flavor scalar meson resonances, the lowest multiplet of tensor resonances can be reasonably understood as standard $\bar{q}q$ mesons within the quark-model picture~\cite{PDG2021}, albeit there are some ongoing disputed discussions about an unusual explanation of the tensor resonance $f_2(1270)$ as a $\rho\rho$ bound state~\cite{Molina:2008jw,Geng:2008gx,Gulmez:2016scm,Du:2018gyn}. The properties of tensor resonances, including the masses and decay widths into two pseudoscalar mesons, are recently computed at unphysically large pion masses by lattice QCD~\cite{Wilson:2014cna,Dudek:2016cru,Briceno:2017qmb}, which can provide us extra useful constraints for the investigation of tensors. In the present lattice simulations~\cite{Wilson:2014cna,Dudek:2016cru,Briceno:2017qmb}, the channels with three or even more light pseudoscalar mesons are not considered.  Likely these multi-particle intermediate states will only affect the existing determination of the masses and two-meson decay widths of the tensor resonances at a moderate level, since the ground tensor resonances are found to be dominated by the compact $q\bar{q}$ components with lightly dressed two-meson decaying states~\cite{Briceno:2017qmb}. The future inclusion of the multi-particle channels will definitely give more information about the multi-particle decay processes of the tensor resonances, where the intermediate vector resonances, such as $\rho,\omega,\phi,K^*$, may play important roles. Nonetheless, in this work we restrict ourselves to the two-body decays of tensor resonances and take into account of the lattice results that also include at most the two-meson intermediate states in the simulations~\cite{Wilson:2014cna,Dudek:2016cru,Briceno:2017qmb}, which still use unphysically large quark masses. A simultaneous analysis of the experimental data and lattice results needs to properly take the extrapolations at different quark masses. Resonance chiral theory ($\rxt$)~\cite{Ecker:1988te} offers a suitable framework to meet this requirement. 
In this work we aim at a precise and comprehensive study of the tensor nonet within $\rxt$. It is hoped that the comprehensive descriptions of the tensor resonance interactions with light pseudoscalar mesons will benefit other related studies involving the tensor resonances, such as the aforementioned processes.

Regarding the mixing of the $SU(3)$ octet and singlet for the light-flavor vector resonances, i.e. the $\omega$-$\phi$ system, the mixing angle $\theta_V=36.3^{\circ}$ only deviates from the ideal mixing scenario $\theta_{\rm id}=35.3^{\circ}$ at the level around 3\%, see the Quark Model review section of Particle Data Group(PDG)~\cite{PDG2021}. In the same place, the tensor-meson mixing angle of the $f_2(1270)$ and $f_2'(1525)$ is determined to be around 30$^\circ$, which is also supported by the previous phenomenological studies~\cite{Cirigliano:2003yq,Giacosa:2005bw}, deviating from the ideal mixing scenario at the level around 15\%, which is roughly at the same level of the $1/N_C$ correction or the $SU(3)$ symmetry breaking effect around 30\%. In the strict ideal-mixing case, the resulting two physical states will contain purely the $\bar{u}u+\bar{d}d$ and $\bar{s}s$ components, whose decay processes will precisely follow the Okubo-Zweig-Iizuka (OZI) rules. In other words, the deviation of the ideal mixing also implies the breaking of OZI rules to a similar extent. The $1/N_C$ expansion in large $N_C$ QCD offers a systematical framework to quantitatively account for the OZI rules~\cite{tHooft:1973alw,Witten:1979kh,Manohar:1998xv}. It is illustrated in Ref.~\cite{Gasser:1984gg} that the $N_C$ counting rule is related with the number of traces in flavor space in chiral perturbation theory ($\chi$PT), although some care needs to be taken in particular situations because of special matrix identities. Generally speaking, one additional trace for a given chiral operator implies one more order of $1/N_C$ suppression. In this way, one can easily impose the OZI rules order by order in the chiral effective field theory. We will follow the R$\chi$T approach~\cite{Ecker:1988te} to include the chiral operators in a joint power counting scheme relying on the simultaneous expansions of momenta, light-quark masses and $1/N_C$, to study the tensor resonance interactions with the light pseudoscalar $\pi, K, \eta$ and $\eta'$, the latter of which corresponds to the pseudo-Nambu-Goldstone boson (pNGB) nonet in the chiral and large $N_C$ limit~\cite{ref:ua1nc}.

The $\eta$-$\eta'$ mixing is another interesting subject in hadron physics, since it is sensitive to both effects from the the light-flavor $SU(3)$ symmetry breaking and the QCD $U_A(1)$ anomaly. Their mixing formalism has been widely investigated in the many phenomenological and lattice analyses. E.g., the $\eta$-$\eta'$ mixing parameters are extensively determined in the radiative decay processes of vector resonances~\cite{Feldmann:1998sh,Escribano:2005qq,Chen:2012vw,Chen:2014yta,Yan:2023nqz}, the eletromagnetic transition form factors involving $\eta$ or $\eta'$~\cite{Escribano:2015nra,Bickert:2020kbn}, the lattice simulations~\cite{Ottnad:2017bjt,Bali:2021qem}, and a recent review in Ref.~\cite{Gan:2020aco}. The various decay processes of tensor resonances can provide another kind of inputs to address the $\eta$-$\eta'$ mixing problem. In this work  different $\eta$-$\eta'$ mixing scenarios will be explored to describe the $T\to PP'$ decays, with $T$ being the tensor resonance. In one of the scenarios, we go beyond the conventional $\eta$-$\eta'$ one-mixing angle formalism by including the two-mixing-angle scheme in the decays of tensor resonances.

The structure of this paper is as follows. The relevant chiral Lagrangians involving tensor resonances and the calculations of their masses and decay widths are discussed in Sec.~\ref{sec.theo}. The tensor masses from experiments and lattice simulations are studied in Sec.~\ref{sec.massfit}, where the $f_2(1270)$-$f_2'(1525)$ mixing is also addressed. In Sec.~\ref{sec.widthfit}, we separately discuss the strong decay processes of $T\to PP'$ and the radiative decay processes of $T\to P\gamma, \gamma\gamma$ in detail. A brief summary and conclusions will be delivered in Sec.~\ref{sec.sum}.

\section{Tensor resonance chiral Lagrangians and calculation of masses and decay amplitudes}\label{sec.theo}

The flavor assignment of the lowest multiplet of tensor-meson resonances  takes the usual form as
\begin{equation}\label{eq.tmatdef}
T_{\mu\nu}=
\begin{pmatrix}
\frac{a^0_2}{\sqrt{2}}+\frac{f^8_2}{\sqrt{6}}+\frac{f^0_2}{\sqrt{3}} &a^+_2&K_2^{*+}\\
a^-_2&-\frac{a^0_2}{\sqrt{2}}+\frac{f^8_2}{\sqrt{6}}+\frac{f^0_2}{\sqrt{3}}&K_2^{*0}\\
K_2^{*-}&\bar K_2^{*0}&-\frac{2f^8_2}{\sqrt{6}}+\frac{f^0_2}{\sqrt{3}}
\end{pmatrix}_{\mu\nu}\,. 
\end{equation} 
We follow Refs.~\cite{Bellucci:1994eb,Ecker:2007us} to construct the chiral Lagrangian involving the tensor-meson resonances with $J^{P}=2^+$, where each of the tensor resonances can be described by the Hermitian symmetric rank-2 tensor field. The kinetic part of Lagrangian reads~\cite{Bellucci:1994eb,Ecker:2007us} 
\begin{align}\label{eq.lagtkin}
\mathscr{L}_{\rm kin}=&-\frac{1}{2}\langle T_{\mu\nu}D^{\mu\nu,\rho\sigma} T_{\rho\sigma}\rangle 
\end{align}
with 
\begin{align}
 D^{\mu\nu,\rho\sigma}=& \big(\Box + M_T^2 \big)\bigg[\frac{1}{2} (g^{\mu\rho}g^{\nu\sigma}+g^{\mu\sigma}g^{\nu\rho}) - g^{\mu\nu}g^{\rho\sigma} \bigg]  \nonumber \\ &
+ g^{\mu\nu}\partial^{\rho}\partial^{\sigma} + g^{\rho\sigma}\partial^{\mu}\partial^{\nu}-\frac{1}{2}\bigg( g^{\mu\sigma}\partial^{\rho}\partial^{\nu}+ g^{\mu\rho}\partial^{\nu}\partial^{\sigma}+g^{\nu\sigma}\partial^{\rho}\partial^{\mu}+ g^{\nu\rho}\partial^{\mu}\partial^{\sigma} \bigg)\,.
\end{align}
For the on-shell tensor resonance, which is indeed the case in our study, it corresponds to a traceless field, i.e. $T^{\mu}_{\,\mu}=0$.  In this case the leading-order (LO) mass term in Eq.~\eqref{eq.lagtkin} reduces to 
\begin{equation} \label{eq.lagmass0}
\mathscr{L}_{m}^{(0)}= -\frac{M_T^2}{2}\langle T_{\mu\nu}  T^{\mu\nu} \rangle \,, 
\end{equation}
which leads to a common mass $M_T$ for all the resonances of the lowest tensor nonet~\eqref{eq.tmatdef}. For the polarization tensor $\epsilon_{\mu\nu}(k;\lambda)$ of a spin-2 tensor resonance with momentum $k$ and spin component $\lambda$, the sum over all of its spin components $\lambda$ gives~\cite{Ecker:2007us} 
\begin{align}\label{eq.sumepsilon}
&\sum_\lambda\epsilon_{\mu\nu}(k;\lambda)\epsilon_{\rho\sigma}^*(k;\lambda)=\frac{1}{2}(P_{\mu\rho} P_{\nu\sigma}+P_{\nu\rho} P_{\mu\sigma})-\frac{1}{3}P_{\mu\nu} P_{\rho\sigma} \,, \quad \bigg( P_{\mu\nu}\equiv g_{\mu\nu}-\frac{k_\mu k_\nu}{M_T^2} \bigg)\,,
\end{align}
which will be used in the calculation of the two-pNGB decay width of the tensor resonance. There are also proposals to treat the tensor resonances as nonrelativistic fields in chiral theory as elaborated in Ref.~\cite{Chow:1997sg}. However we will stick to the realistic description in this work. 

Higher-order chiral operators are required to implement the mass splittings. As mentioned in the Introduction, both the light-quark masses and $1/N_C$ corrections can be relevant to achieve the realistic description for the tensor nonet.  Therefore, the joint expansion scheme relying on the simultaneous expansion of momenta, light-quark masses and $1/N_C$, will be useful to  organize the pertinent chiral Lagrangians involving the tensor resonances. The general structure of the operator in $\rxt$ can be written as
\begin{eqnarray}
 \langle R_1 R_2 \cdots R_j \chi^{(n)}(\Phi) \rangle\,, 
\end{eqnarray}
where insertions of additional traces should be understood,  $R_{i=1,2,...,j}$ stand for the resonance fields, and $\chi^{(n)}(\Phi)$ denotes the chiral building block consisting of the pNGBs $\Phi$ with the joint-expansion counting index $n$. At next-to-leading order (NLO), the relevant Lagrangian to the mass terms consist of two operators 
\begin{eqnarray}\label{eq.lagmass1}
\mathscr{L}_{m}^{(1)} = \lambda_T\langle T_{\mu\nu}T^{\mu\nu}\chi_+\rangle +  \lambda_T^\prime \langle T_{\mu\nu}\rangle\langle T^{\mu\nu} \rangle \,,
\end{eqnarray}
where the first and second terms scale as $O(p^2, N_C^{0})$ and $O(p^0,N_C^{-1})$, respectively, compared to the LO operator in Eq.~\eqref{eq.lagmass0}. According to the R$\chi$T formalism~\cite{Ecker:1988te}, the LO tensor mass $M_T$ scales as $O(p^0,N_C^{0})$, consistent with the behavior of a $q\bar{q}$ meson at large $N_C$~\cite{tHooft:1973alw,Witten:1979kh,Manohar:1998xv}. In other words, the operator accompanied by $\lambda_T$ respects the OZI rule, while the $\lambda_T^\prime$ term violates this rule. I.e., the deviation of the ideal mixing in the $f_2^8$-$f_2^0$ system is contributed by the $\lambda_T^\prime$ term. One could also try to introduce the relevant Lagrangian to the mass terms at next-to-next-to-leading order (NNLO), which would contain three additional operators: $\langle T_{\mu\nu}T^{\mu\nu}\chi_+\chi_+\rangle$, $\langle T_{\mu\nu}T^{\mu\nu}\rangle \langle \chi_+ \rangle$, $\langle T_{\mu\nu} \rangle \langle T^{\mu\nu} \chi_+ \rangle$. By focusing on the mass splittings along, it is impossible to pin down the NNLO coefficients accompanying these operators. 
It is noted that the $\langle T_{\mu\nu} \rangle \langle T^{\mu\nu} \chi_+ \rangle$ operator was included together with the NLO ones~\eqref{eq.lagmass1} in Ref.~\cite{Cirigliano:2003yq} to address the tensor masses. While in this work, we will try to follow as much as the joint power counting scheme by relying on the simultaneous expansions of light-quark masses and $1/N_C$ at NLO.  

For the two-pNGB decay processes of the tensor resonances, the LO $\rxt$ Lagrangian comprises one operator 
\begin{align}\label{eq.lagint0}
\mathscr{L}_{TPP}^{(0)}=  g_{T}\langle T_{\mu\nu}\{u^\mu,u^\nu\}\rangle\,,
\end{align}
and the relevant NLO Lagrangian consists of four terms 
\begin{align}\label{eq.lagint1}
\mathscr{L}_{TPP}^{(1)}=&  f_{T}\langle T_{\mu\nu}\{\{u^\mu,u^\nu\},\chi_{+}\}\rangle+f_{T}^\prime\langle
T_{\mu\nu}(u^\mu\chi_{+}u^\nu+u^\nu\chi_{+}u^\mu)\rangle \nonumber \\ & 
+g_{T}^\prime\langle T_{\mu\nu}\rangle\langle u^\mu u^\nu \rangle +g_{T}^{\prime\prime}\big( \langle T_{\mu\nu} u^\mu\rangle\langle  u^\nu \rangle + \langle T_{\mu\nu} u^\nu\rangle\langle  u^\mu \rangle \big)\,, 
\end{align}
with the basic chiral building blocks 
\begin{align}
&u=e^{i\frac{\Phi}{\sqrt{2}F}}\,, \quad u_{\mu}=i\{u^{\dagger}\partial_\mu u-u\partial_\mu u^{\dagger}\}\,, \quad
\chi_{+}=u^{\dagger} \chi u^{\dagger}+u\chi^{\dagger}u\,,\quad \chi=2B(s+ip),
\quad 
\end{align}
and the pNGB matrix 
\begin{equation}
\Phi=
\begin{pmatrix}
\frac{\pi^0}{\sqrt{2}}+\frac{\eta^8}{\sqrt{6}}+\frac{\eta^0}{\sqrt{3}}&\pi^+&K^{+}\\
\pi^-&-\frac{\pi^0}{\sqrt{2}}+\frac{\eta^8}{\sqrt{6}}+\frac{\eta^0}{\sqrt{3}}&K^{0}\\
K^{-}&\bar K^{0}&-\frac{2\eta^8}{\sqrt{6}}+\frac{\eta^0}{\sqrt{3}}
\end{pmatrix}. 
\end{equation}
The light quark masses are introduced by taking the external scalar source as $s={\rm diag}(m_u,m_d,m_s)$. We will take the isospin symmetric limit, i.e. $m_u=m_d$, throughout this work. Compared to the LO operator $g_T$ in Eq.~\eqref{eq.lagint0} that scales as $O(p^2,N_C^1)$, the NLO operators accompanied by $f_T$ and $f_T^\prime$ in Eq.~\eqref{eq.lagint1} are suppressed by the factor of $O(p^2,N_C^0)$, while the other two terms of $g_T^\prime$ and $g_T^{\prime\prime}$ are suppressed as $O(p^0,N_C^{-1})$. Two additional operators $\beta g^{\mu\nu} \langle T_{\mu\nu} u_\rho u^\rho \rangle$ and $\gamma g^{\mu\nu}\langle T_{\mu\nu} \chi_+ \rangle $ are introduced in Ref.~\cite{Ecker:2007us} to describe the interactions of light pseudoscalar mesons and tensor resonances. Nevertheless, these two additional terms do not contribute to the decay widths of the tensor resonances, since the latter are traceless on shell.

The NLO operators in Eq.~\eqref{eq.lagmass1} not only introduce the  mass splittings but also lead to the mass mixing of the octet $f_2^8$ and singlet $f_2^0$. The physical tensor resonances $f_2$ and $f_2'$ correspond to the mass eigenstates after performing diagonalization of the octet and singlet fields. We parametrize the mixing relation between the octet-singlet bases and the physical ones as
\begin{align}\label{eq.mixf2}
f_2^8=\sin{\theta_T} f_2+\cos{\theta_T} f_2'\,,\notag\\
f_2^0=\cos{\theta_T} f_2-\sin{\theta_T} f_2'\,,
\end{align}
with $\theta_T$ the tensor mixing angle. In the ideal mixing scenario, one has $\theta_T=35.3^\circ$. The masses of the physical resonances after the diagonalization procedure are found to be 
\begin{align}\label{eq.massf2}
M^2_{f_2}= M_T^2 - 4\lambda_T m_K^2 - 3\lambda_T^\prime- \sqrt{ 16\lambda_T^2 (m_K^2-m_\pi^2)^2 - 8\lambda_T \lambda_T^\prime (m_K^2-m_\pi^2) + 9 {\lambda_T^\prime}^2  }\,, 
\end{align}
\begin{align}
M^2_{f_2'}= M_T^2 - 4\lambda_T m_K^2 - 3\lambda_T^\prime + \sqrt{ 16\lambda_T^2 (m_K^2-m_\pi^2)^2 - 8\lambda_T \lambda_T^\prime  (m_K^2-m_\pi^2) + 9 {\lambda_T^\prime}^2  }\,, 
\end{align}
and the masses of $a_2$ and $K_2^*$ read 
\begin{align}
M_{a_2}^2=M_T^2-4\lambda_Tm_\pi^2\,,
\end{align}
\begin{align}
M_{K_2^*}^2=M_T^2-4\lambda_Tm_K^2\,. \label{eq.masskv2}
\end{align}
For the tensor-meson mixing angle $\theta_T$, we have 
\begin{align}\label{eq.deftan2t}
\tan{2\theta_T}=\frac{8\sqrt{2}(m_\pi^2-m_K^2)\lambda_T}
{4( m_\pi^2-m_K^2)\lambda_T+9\lambda_T^\prime}\,,
\end{align} 
which recovers the ideal mixing result, i.e., $\theta_T=\arctan{(2\sqrt{2})}/2=35.3^\circ$, by setting the $1/N_C$ suppressed coupling $\lambda_T^\prime=0$, as expected.

Next we calculate the two-pNGB decay widths of the tensor resonances. Another prerequisite is to solve the $\eta$-$\eta'$ mixing, which is a more involved system than the $f_2$-$f_2'$ case. A modern study relying on $U(3)$ chiral perturbation theory~\cite{Leutwyler:1997yr,Kaiser:2000gs} for the $\eta$-$\eta'$ mixing gives the two-angle-mixing prescription 
\begin{equation}\label{eq.mixeta08}
\left( \begin{array}{ccc}
\eta  \\ \eta'    \\
\end{array} \right)\,
=\frac{1}{F} \left( \begin{array}{ccc}
F_8 \cos\theta_8 & -F_0 \sin\theta_0  \\
F_8 \sin\theta_8 & F_0 \cos\theta_0   \\
\end{array} \right)\,
\left( \begin{array}{ccc}
\eta_8  \\ \eta_0    \\
\end{array} \right)\,,
\end{equation}
which naturally reduces to the one-mixing-angle case by taking $F_8=F_0=F$ and $\theta_8=\theta_0=\theta$.

By using the interaction Lagrangians~\eqref{eq.lagint0}~\eqref{eq.lagint1}, the mixing formulas in Eqs.~\eqref{eq.mixf2}~\eqref{eq.mixeta08} and the sum of polarization tensors~\eqref{eq.sumepsilon}, it is then straightforward to calculate the two-pNGB decay width for the tensor resonance. We give the expressions for all the relevant $T\to P P'$ decay widths in the Appendix.

\section{Mass splittings and tensor mixing angle}\label{sec.massfit}

In this part, we will rely on the LO~\eqref{eq.lagmass0} and NLO~\eqref{eq.lagmass1} Lagrangians, i.e., the results in Eqs.~\eqref{eq.massf2}-\eqref{eq.masskv2}, to analyze the masses of $f_2$, $f_2'$, $a_2$ and $K_2^*$. As a byproduct, we can also predict the $f_2$-$f_2'$ mixing angle $\theta_T$ through the formula in Eq.~\eqref{eq.deftan2t}. For the inputs of experimental masses, we will take the PDG averages~\cite{PDG2021}. Since the isospin breaking effects are neglected in our study, we take the average mass between the neutral and charged $K_2^*(1430)$ and also assign a conservative uncertainty to cover both the neutral and charged masses. The experimental (Exp) inputs for the masses used in our fits read  
\begin{align}
M_{f_2}^{\rm Exp}= 1275.5\pm 0.8\,, \quad M_{a_2}^{\rm Exp}= 1318.2\pm 0.6\,, \quad M_{K_2^*}^{\rm Exp}= 1429.9\pm 4.1\,, \quad  M_{f_2'}^{\rm Exp}= 1517.4\pm 2.5\,, 
\end{align}
which are given in units of MeV. 

The three parameters relevant to the tensor masses determined in the fits to experimental data are 
\begin{align}\label{eq.parafitexp}
M_T=(1308.5\pm 1.2)~{\rm MeV}\,,\quad  \lambda_T= -0.336\pm 0.008\,, \quad 
\lambda_T'= (25718\pm 1054)~{\rm MeV^2}\,,  
\end{align}
with $\chi^2/(d.o.f)= 0.05/(4-3)$. According to the almost vanishing $\chi^2$, one can safely conclude that the NLO expressions of $\rxt$ in Eqs.~\eqref{eq.massf2}-\eqref{eq.masskv2} are sufficient to describe the masses of the lowest tensor nonet. The prediction of the tensor mixing angle $\theta_T$ via Eq.~\eqref{eq.deftan2t} at the physical masses is 
\begin{equation}
\theta_T^{\rm Phy} = (29.1 \pm 0.1)^\circ\,,
\end{equation}
which is close to the previous determinations in Refs.~\cite{Cirigliano:2003yq,Giacosa:2005bw,Chow:1997sg}.

Recent lattice simulations on the unstable resonances have made noticeable progresses. E.g., both the masses and partial decay widths of the lowest multiplet of tensor resonances have been determined at unphysically large pion mass in Refs.~\cite{Wilson:2014cna,Dudek:2016cru,Briceno:2017qmb}. The explicit values of the masses from the lattice determinations read 
\begin{align}\label{eq.massf2lat}
M_{f_2}^{\rm Lat}= 1470\pm 15\,, \quad M_{a_2}^{\rm Lat}= 1505\pm 5\,, \quad M_{K_2^*}^{\rm Lat}= 1577\pm 7\,, \quad  M_{f_2'}^{\rm Lat}= 1602\pm 10\,, 
\end{align}
which are given in units of MeV. The corresponding masses of $\pi, K,\eta$, and $\eta'$ used in the lattice simulations~\cite{Briceno:2017qmb} are 
\begin{align}\label{eq.massmpilat}
m_\pi^{\rm Lat}= 391 \pm 1\,, \quad m_K^{\rm Lat}= 550\pm 1\,, \quad m_{\eta}^{\rm Lat}= 587\pm 1\,, \quad  m_{\eta'}^{\rm Lat}= 930\pm 6\,, 
\end{align}
which are in units of MeV as well. Utilizing the NLO expressions  ~\eqref{eq.massf2}-\eqref{eq.masskv2} to fit the lattice results in Eq.~\eqref{eq.massf2lat} with the lattice $m_\pi$ and $m_K$ in Eq.~\eqref{eq.massmpilat}, the three parameters entering in the tensor masses are found to be 
\begin{align} \label{eq.parafitlat}
M_T=(1444 \pm 18)~{\rm MeV}\,,\quad  \lambda_T= -0.307\pm 0.052\,, \quad 
\lambda_T'=( 27920\pm 16526 )~{ \rm MeV^2}\,,  
\end{align}
with $\chi^2/(d.o.f)= 3.3/(4-3)$. The corresponding masses from such fit are 
\begin{align}
&M_{f_2}^{\rm Lat,Fit}= 1465 \pm {26}\,, \quad M_{a_2}^{\rm Lat,Fit}= 1508 \pm{9}\,, \quad \nonumber\\ &  M_{K_2^*}^{\rm Lat,Fit}= 1568\pm {9}\,,  \quad  M_{f_2'}^{\rm Lat,Fit}= 1612\pm {15}\,, \label{eq.massf2lattheo}
\end{align} 
which reasonably reproduce the lattice inputs~\eqref{eq.massf2lat}, though with somewhat large $\chi^2$. The tensor mixing angle evaluated at the unphysically large meson masses in Eq.~\eqref{eq.massmpilat} is 
\begin{equation}
\theta_T^{\rm Lat} = (25.0 {^{+5.6}_{-4.4}})^\circ\,. 
\end{equation}

By further comparing the two sets of parameters in Eqs.~\eqref{eq.parafitexp} and \eqref{eq.parafitlat} that are determined from the fits to experiment and lattice data, respectively, we find that the two NLO couplings $\lambda_T$ and $\lambda_T'$ from the two fits are compatible within uncertainties, while the two sets of data clearly require different values for the LO mass $M_T$. In other words, the relative mass splittings among the lowest tensor multiplet both from the experiment and lattice determinations can be well described by the two operators in Eq.~\eqref{eq.lagmass1}, although one needs to globally shift the LO mass $M_T$ when separately fitting the experimental and lattice data. One way to understand this shift is to resort to a specific NNLO operator 
\begin{equation}\label{eq.lagmass2}
\lambda_T^{\prime\prime}\langle T_{\mu\nu} T^{\mu\nu} \rangle \langle \chi_+ \rangle\,, 
\end{equation}
since this operator will contribute to a common mass $-4\lambda_T^{\prime\prime}(2m_K^2+m_\pi^2)$ for all the tensor nonet, namely it is equivalent to redefine $M_T^2$ as $M_T^2-4\lambda_T^{\prime\prime}(2m_K^2+m_\pi^2)$. It is pointed out that the $\lambda_T^{\prime\prime}$ term will not affect the expression of tensor mixing angle given in Eq.~\eqref{eq.deftan2t}. As a result, we can simultaneously describe both experiment and lattice determinations of the masses for the lowest tensor multiplet by using the LO $M_T$~\eqref{eq.lagmass0}, the NLO $\lambda_T$ and $\lambda_T^\prime$~\eqref{eq.lagmass1}, and the NNLO $\lambda_T^{\prime\prime}$~\eqref{eq.lagmass2}. The corresponding parameters from the joint fit to experimental and lattice masses   are 
\begin{align}\label{eq.parafitexplat}
& M_T=(998.7\pm 26.4)~{\rm MeV}\,,\qquad   \lambda_T= -0.335\pm 0.009\,,   \nonumber\\ &
\lambda_T'=( 25732\pm 1216)~{\rm MeV^2} \,,\,\quad  \lambda_T^{\prime\prime}= -0.350\pm 0.026\,,
\end{align}
with $\chi^2/(d.o.f)=4.5/(8-4)$. The resulting values for the tensor masses and the tensor mixing angles at the physical and lattice meson masses from the joint fit, together with the two types of inputs, are summarized in Table~\ref{tab.mass}. The tensor mixing angle evaluated at unphysically large meson masses~\eqref{eq.massmpilat} is slightly decreased, when comparing with its physical value.

\begin{table}[ht]
\centering
\begin{threeparttable}
\begin{tabular}{c|cc|cc}
\hline
                     &       Exp               &    Theo    & Lat  &  Theo                  \\  \hline
   $M_{f_2}({\rm MeV})$     &  $1275.5\pm0.8$    &   $1275.5{\pm 1.7}$ &   $1470 \pm 15$  & $1466 {\pm 8}$            \\ 
$M_{a_2} ({\rm MeV})$       &  $1318.2\pm0.6$  & $1318.2 {\pm 1.3}$ &  $1505\pm 5$ &  $1505 {\pm 8}$              \\ 
$M_{K_2^*}({\rm MeV})$&  $1429.9\pm4.1$   &  $1428.8 {\pm 2.8}$ &    $1577 \pm 7$   & $1570 {\pm 8}$                 \\ 
$M_{f_2^\prime}({\rm MeV})$& $1517.4\pm2.5$ & $1517.2 {\pm 5.1}$ &$1602\pm10$& $1620 {\pm 8}$ \\
$\theta_T\,(^\circ)$ & $--$ & $29.0{\pm 0.4}$ &$--$ & $26.4{\pm 0.3}$ \\  \hline
\end{tabular}
\end{threeparttable}
\caption{The results from the joint fit (the column labeled as Theo) to the experimental (Exp)~\cite{PDG2021} and lattice (Lat)~\cite{Briceno:2017qmb} determinations of the tensor masses. The predictions to the tensor mixing angles are obtained according to the formula in Eq.~\eqref{eq.deftan2t}.  }\label{tab.mass}
\end{table}

\begin{figure}[htbp]
\centering
\includegraphics[width=0.45\textwidth,angle=-0]{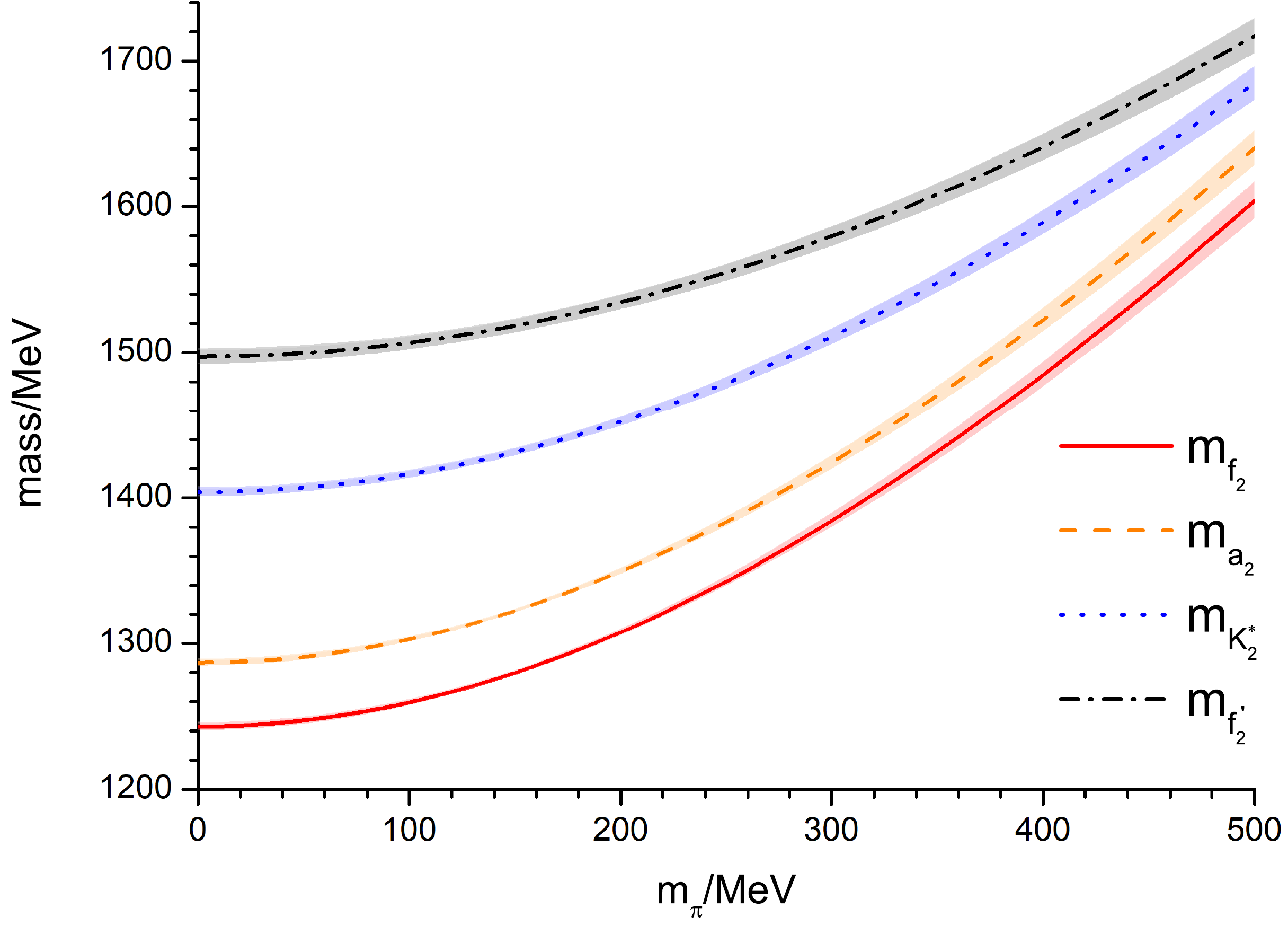} 
\includegraphics[width=0.45\textwidth,angle=-0]{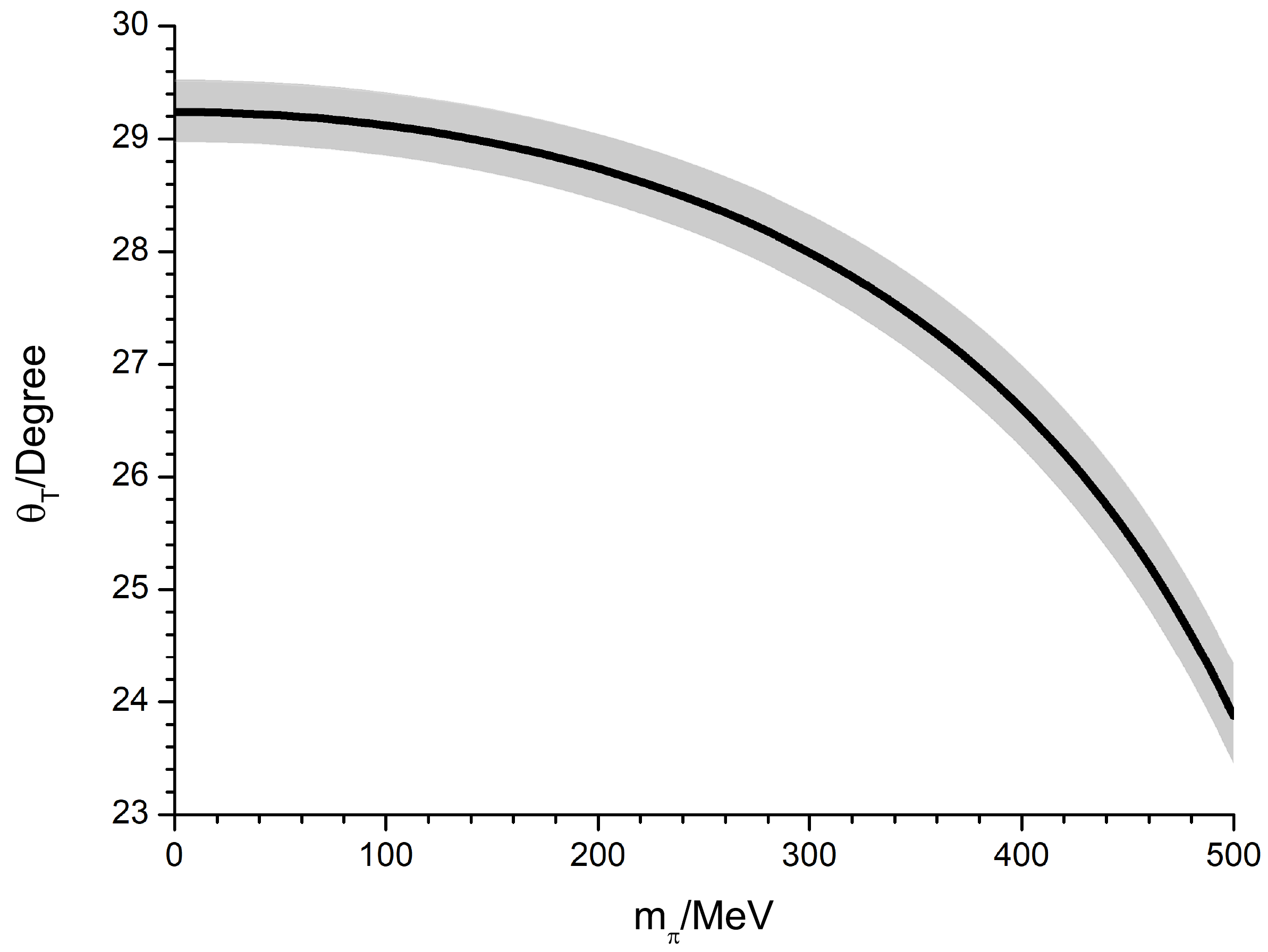} 
\caption{Masses (left panel) and mixing angle (right panel) of the tensor resonances as functions of $m_\pi$. The mass of strange quark is set at its physical value and the shaded areas correspond to our estimation of the theoretical uncertainties, see the text for details.}
 \label{fig.massandtheta}
\end{figure}

Relying on the resonance parameters in Eq.~\eqref{eq.parafitexplat}, we give predictions to the tensor masses and their mixing angles in Fig.~\ref{fig.massandtheta} at different pion masses, ranging from chiral limit to $m_\pi=500$~MeV. Since nowadays most of the lattice calculations can run   simulations by using the physical value for the strange quark mass, we perform the chiral extrapolations in Fig.~\ref{fig.massandtheta} by fixing the strange quark mass at its physical value. To be more specific, the leading-order relation $m_{K,{\rm Lat}}^2=m_{K,{\rm Phy}}^2 + \big(m_{\pi,{\rm Lat}}^2-m_{\pi,{\rm Phy}}^2\big)/2$~\cite{Gasser:1984gg} has been employed to extrapolate the kaon masses at different pion masses. We have also explicitly verified that the results are barely changed by taking into account the more sophisticated higher-order chiral corrections for $m_K^2$ in Refs.~\cite{Guo:2015xva,Gu:2018swy,Gao:2022xqz}.
The shaded areas correspond to the theoretical uncertainties that are estimated by taking the parameters in Eq.~\eqref{eq.parafitexplat}. 

\section{Decay widths of the tensor resonances}\label{sec.widthfit}

In this section, we first discuss the various $T\to P P'$ decay widths relying on the $\rxt$ Lagrangians in Eqs.~\eqref{eq.lagint0} and \eqref{eq.lagint1}. The explicit expressions of the decay widths can be found in the Appendix. 
We will simultaneously include the $T\to P P'$ decay widths both from the experiments and lattice study as inputs in the joint fit. 

For the inputs of $T \to P P'$ partial decay widths from the experiments, we will take the PDG averages from Ref.~\cite{PDG2021}. All the available two-pseudoscalar-meson decay widths of $f_2, f_2', a_2$ and $K_2^*$ from PDG~\cite{PDG2021} are given in Table~\ref{tab.physwidth}. 
Although the explicit values for different partial decay widths of tensor resonances into the two pseudoscalar mesons are not provided in the lattice analysis~\cite{Briceno:2017qmb}, they can be inferred from either the available branching ratios for $f_2^{(')}\to \pi\pi, K\bar{K}$~\cite{Briceno:2017qmb} or the coupling strengths to different channels for the cases of $a_2\to \pi\eta, K\bar{K}$~\cite{Dudek:2016cru}. Since the width of $K_2^*\to K\eta$ at the large meson masses~\eqref{eq.massmpilat} seems very small~\cite{Wilson:2014cna}, we will estimate the decay width of $K_2^*\to K\pi$ by the pole width provided by the former reference. 
The accessible decay widths from the Hadron Spectrum Collaboration(HSC) lattice simulations~\cite{Wilson:2014cna,Briceno:2017qmb} are summarized in Table~\ref{tab.latwidth}. 

To describe the processes involving $\eta$ and $\eta'$, one needs to introduce their mixing parameters as shown in Eq.~\eqref{eq.mixeta08}. In the following, we will distinguish two scenarios for the $\eta$-$\eta'$ mixing to proceed the discussions. In the first scenario, the conventional one-mixing-angle formalism will be used to describe the experimental and lattice data. In the other scenario, the two-mixing-angle formula will be employed to study the two-pseudoscalar-meson decay widths from the experiments and lattice simulations. We separately discuss the two scenarios in the following.  

\subsection{ $T\to P P'$ decay widths with one $\eta$-$\eta'$ mixing angle}

We point out a subtlety that needs to be taken care of in the simultaneous description of the experimental data and lattice decay widths at unphysically meson masses. Apart from the explicit light-pseudoscalar-meson mass dependence of the decay widths as shown in the expressions in Appendix, the meson-mass dependence of the mixing angles for both the $f_2$-$f_2'$ and $\eta$-$\eta'$ systems should be considered as well. Regarding the former case, we will take the values of mixing angle $\theta_T$ determined in the analysis of tensor masses, as shown in Table~\ref{tab.mass}.  

For the $\eta$-$\eta'$ case, due to the QCD $U_A(1)$ anomaly, its mixing mechanism is more involved and the two-mixing-angle scheme as proposed in Ref.~\cite{Leutwyler:1997yr} naturally results after including the higher order $U(3)$ $\chi$PT effects~\cite{Guo:2015xva,Gu:2018swy,Gao:2022xqz}. 
While, the one-mixing-angle description for $\eta$-$\eta'$ naturally results from the LO $\chi$PT, since there is only the mass mixing term at this order. The $T\to P P'$ decay widths in the one-mixing-angle case can be obtained by taking $\theta_0=\theta_8=\theta$ and $F_0=F_8=F$ in the expressions given in the Appendix. In accord with this LO treatment of the $\eta$-$\eta'$ system, we also adopt in the one-mixing-angle scenario the LO result for the weak decay constants of $\pi$ and $K$, i.e., by taking the common weak decay constant $F$ at chiral limit for $\pi$ and $K$ in the decay amplitudes given in the Appendix. To be more specific, we will take the value of $F=81.7$~MeV determined in Ref.~\cite{Guo:2015xva}. 
For the sake of completeness, we give the expression for the LO $\eta$-$\eta'$ mixing angle~\cite{Guo:2011pa}
\begin{align} \label{eq.mixetalo}
\sin{\theta} = -\left( \sqrt{1 +
\frac{ \big(3M_0^2 - 2\Delta^2 +\sqrt{9M_0^4-12 M_0^2 \Delta^2 +36 \Delta^4 } \big)^2}{32 \Delta^4} } ~\right )^{-1}\,,
\end{align}
where $\Delta^2=m_K^2 - m_\pi^2$ and $M_0$ corresponds to the mass of the singlet $\eta_0$ in the chiral limit. The above equation offers a viable approach to extrapolate the $\eta$-$\eta'$ mixing angle from the physical masses to unphysical lattice masses.

\begin{table}[!htb]
	\centering
	\begin{tabular}{cccc}
		\hline\hline
&Exp~\cite{PDG2021}  & Fit-I & Fit-II \\ \hline
$\Gamma_{f_2\rightarrow\pi\pi}$        &$157.2\pm7.3$      & $156.9\pm 14.3$  & $157.2\pm 15.5$\\ \hline
$\Gamma_{f_2\rightarrow K\bar{K}}$&$8.6\pm1.1$        & $9.4\pm 1.5$     & $8.6\pm 1.6$  \\ \hline
$\Gamma_{f_2\rightarrow\eta\eta}$      & $0.7\pm0.2$       & $1.0\pm 0.2$     & $0.8\pm 0.3$\\ \hline
$\Gamma_{f_2^\prime\rightarrow\pi\pi}$    & $0.7\pm0.2$       & $0.6\pm 0.5$     & $0.7\pm 0.4$\\ \hline
$\Gamma_{f_2^\prime\rightarrow K\bar{K}}$ &$75.3\pm6.3$        & $74.1\pm 12.4$   & $70.2\pm 12.3$  \\ \hline
$\Gamma_{f_2^\prime\rightarrow\eta\eta}$  & $10.0\pm2.6$      & $9.7\pm 3.5$     & $7.5\pm 2.9$\\ \hline
$\Gamma_{a_2\rightarrow K\bar{K}}$& $5.2\pm1.1$       & $3.3\pm 1.2$     & $4.0\pm 1.3$\\ \hline
$\Gamma_{a_2\rightarrow\pi\eta}$       & $15.5\pm2.1$      & $13.3\pm 2.9$    & $13.2\pm 3.4$ \\ \hline
$ \Gamma_{a_2\rightarrow\pi\eta^\prime}$  & $0.6\pm0.1$       & $0.6\pm 0.1$     & $0.6\pm 0.1$  \\ \hline
$\Gamma_{K_2^*\rightarrow \eta K} $    &$0.2^{+0.4}_{-0.2}$& 0.2$^{+0.4}_{-0.2}$& $0.1^{+0.4}_{-0.1}$   \\ \hline
$\Gamma_{K_2^*\rightarrow\pi K}$       & $52.1\pm6.1$      & $53.5\pm 6.4$     & $59.9\pm 7.1$ \\ \hline\hline 
	\end{tabular}
	\caption{ Partial decay widths of the tensor resonances at physical meson masses. The columns labeled as Fit-I and Fit-II correspond to the fit  results from the one-mixing-angle and two-mixing-angle scenarios, respectively, see the text for details. All the numbers are given in units of MeV.  }\label{tab.physwidth}
\end{table}

\begin{table}[!hbt]
	\centering
	\begin{tabular}{cccc}
		\hline\hline
 &Lat~\cite{Briceno:2017qmb,Wilson:2014cna}  & Fit-I & Fit-II \\ \hline
$\Gamma_{f_2\rightarrow\pi\pi}$       &$136.0\pm22.0$&$132.0\pm 10.8$          & $117.0\pm 10.2$  \\ \hline
$\Gamma_{f_2\rightarrow K\bar{K}}$    &$19.2\pm7.8$  &$24.4\pm 3.7$            & $20.9\pm 3.3$\\ \hline
$\Gamma_{f_2^\prime\rightarrow\pi\pi}$   &$4.3\pm3.1$   &$(1.2^{+16.5}_{-1.2})\times10^{-2}$ & $0.2\pm 0.2$  \\ \hline
$\Gamma_{f_2^\prime\rightarrow K\bar{K}}$&$49.7\pm20.1$ &$59.7\pm 9.2$            & $53.4\pm 7.8$  \\ \hline
$\Gamma_{a_2\rightarrow K\bar{K}}$    &$7.1\pm1.9$   &$8.0\pm 2.3$             & $9.2\pm 2.2$ \\ \hline
$\Gamma_{a_2\rightarrow\pi\eta}$      &$13.1\pm3.0$  &$13.8\pm 2.5$            & $17.9\pm 2.0$  \\ \hline
$\Gamma_{K_2^*\rightarrow\pi K}$      &${62\pm 12}$&$47.4\pm 6.3$ & $48.4\pm 6.4$   \\ \hline\hline
	\end{tabular}
	\caption{ Partial decay widths of the tensor resonances with lattice  masses~\eqref{eq.massmpilat} and \eqref{eq.massf2lat}. The meanings of Fit-I and Fit-II are the same as Table~\ref{tab.physwidth}. All the numbers are given in units of MeV. }\label{tab.latwidth}
\end{table}

Now we are ready to perform the joint fit to the $T\to P P'$ decay widths both from experiments and lattice simulations in the one-mixing-angle scenario. The resulting parameters from the joint fit are 
\begin{eqnarray}\label{eq.fiti}
&g_T= (16.6\pm 1.1)~{\rm MeV}\,, \quad &f_T= (3.9\pm 1.4)\times 10^{-6}~{\rm MeV^{-1}}\,, \nonumber \\
&g'_T= (4.8\pm 0.6)~{\rm MeV}\,, \quad &f'_T= (-3.1\pm 2.9)\times 10^{-6}~{\rm MeV^{-1}}\,,\quad \theta^{\rm Phy}= (-9.0\pm 5.5)^\circ \,,  \nonumber \\
\end{eqnarray} 
with $\chi^2/({\rm d.o.f})= 11.4/(18-5)$. 
 The various partial decay widths of the tensor resonances from the one-mixing-angle fit are collected in the column labeled as Fit-I in Tables~\ref{tab.physwidth} and \ref{tab.latwidth} for the physical and lattice masses, respectively, together with the experimental and lattice inputs. Both the decay widths from the experiments and lattice simulations are reasonably reproduced in our fits. 
For the tensor mixing angle $\theta_T$ at physical and lattice masses~\eqref{eq.massmpilat}, we have fixed their values from the analysis of tensor masses in Table~\ref{tab.mass}. 
According to Eq.~\eqref{eq.mixetalo}, the $\eta$-$\eta'$ mixing angle $\theta^{\rm Phy}=-(9.0\pm 5.5)^{\circ}$ at physical masses corresponds to the solution of $M_0=(1210^{+600}_{-250})$~MeV, which is clearly larger than the determinations in the $U(3)$ $\chi$PT~\cite{Guo:2015xva,Gu:2018swy,Gao:2022xqz}. 
In other words, although the one-mixing-angle fit can well reproduce the experimental and lattice decay widths for the various $T\to P P'$ processes, there is a tension between the resulting LO $\eta$-$\eta'$ mixing angle $\theta$ from such fit and the one from the determination of $U(3)$ $\chi$PT~\cite{Guo:2015xva,Gu:2018swy,Gao:2022xqz}. 
Therefore it is interesting to pursue the two-mixing-angle formalism to fit $T\to P P'$ decay widths.

\begin{figure}[htbp]
\centering
\includegraphics[width=0.49\textwidth,angle=-0]{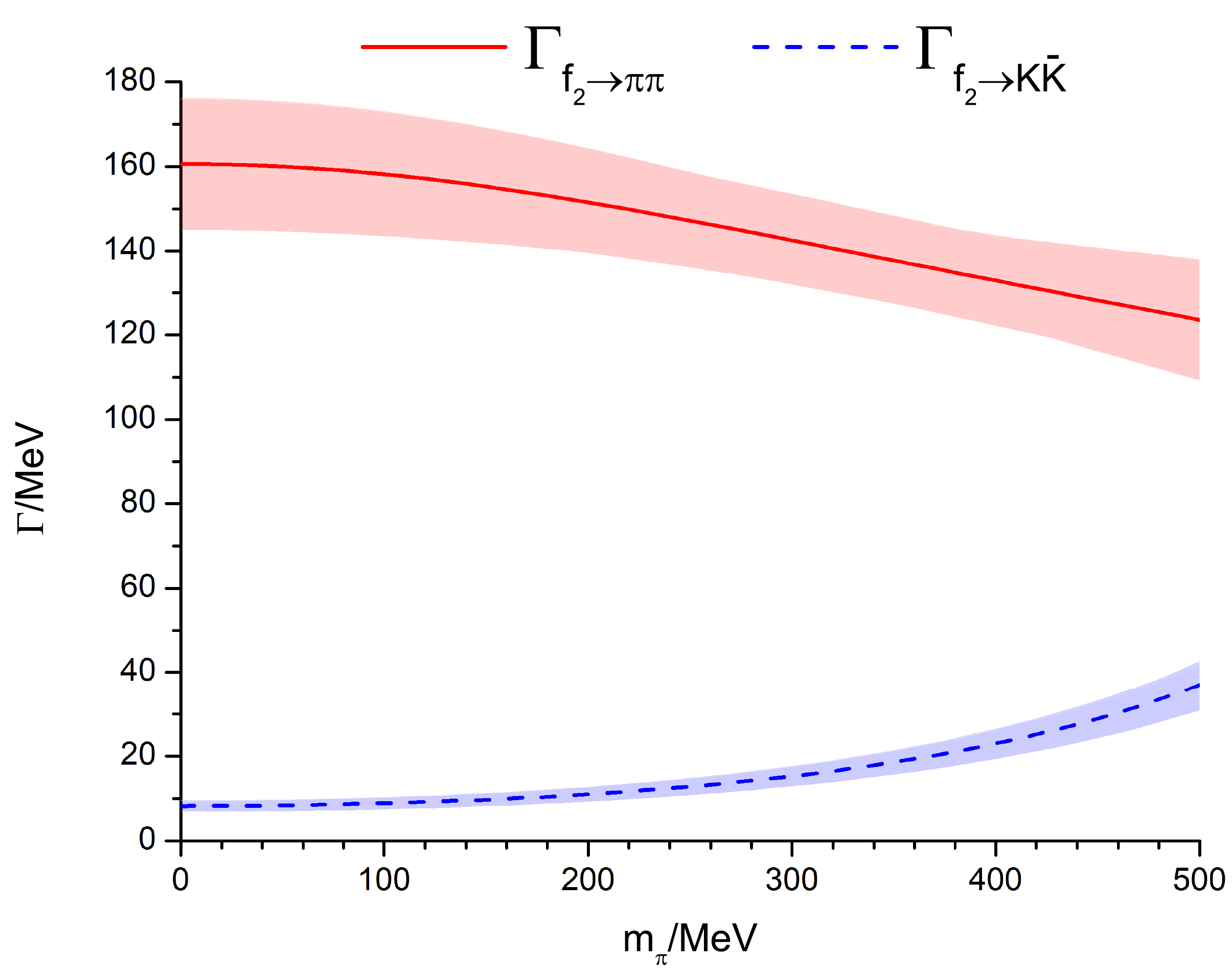}
\includegraphics[width=0.49\textwidth,angle=-0]{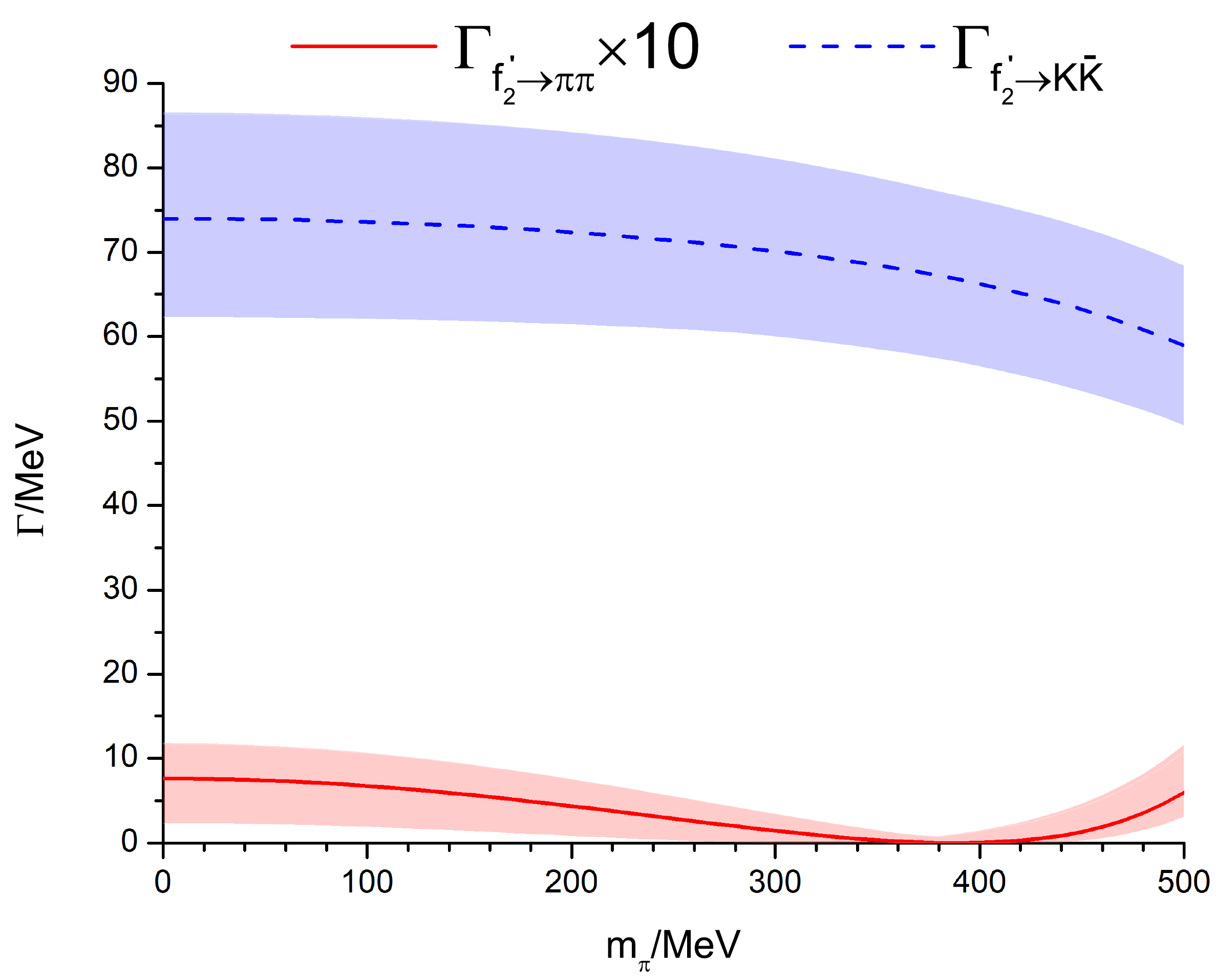}
\caption{Partial two-meson decay widths of $f_2$ (left panel) and $f_2'$  (right panel) as functions of $m_\pi$. The mass of strange quark is set at its physical value and the shaded areas correspond to our estimation of the theoretical uncertainties, see the text for details.   }
 \label{fig.widthf2}
\end{figure}

\begin{figure}[htbp]
 \centering
\includegraphics[width=0.49\textwidth,angle=-0]{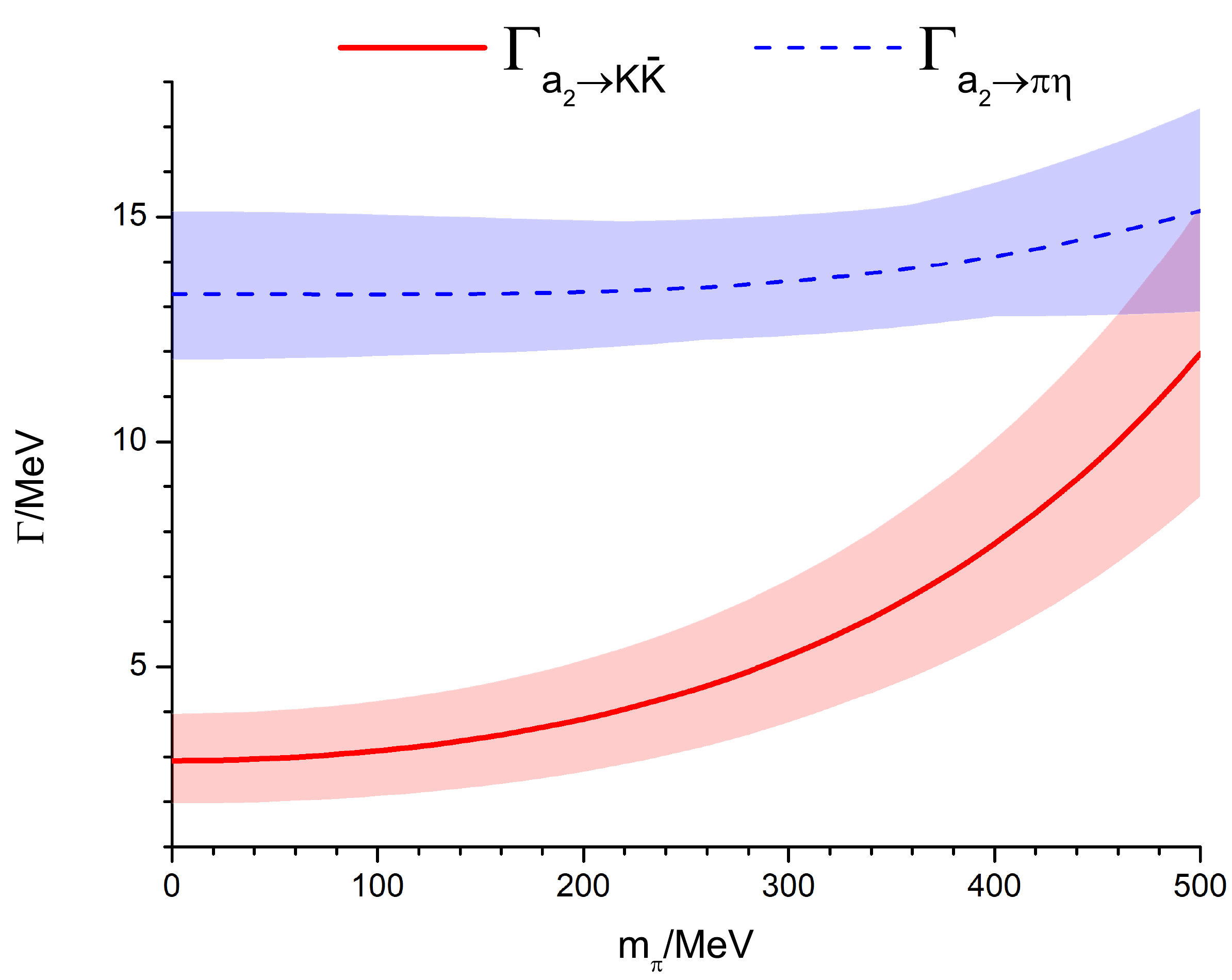} 
\includegraphics[width=0.49\textwidth,angle=-0]{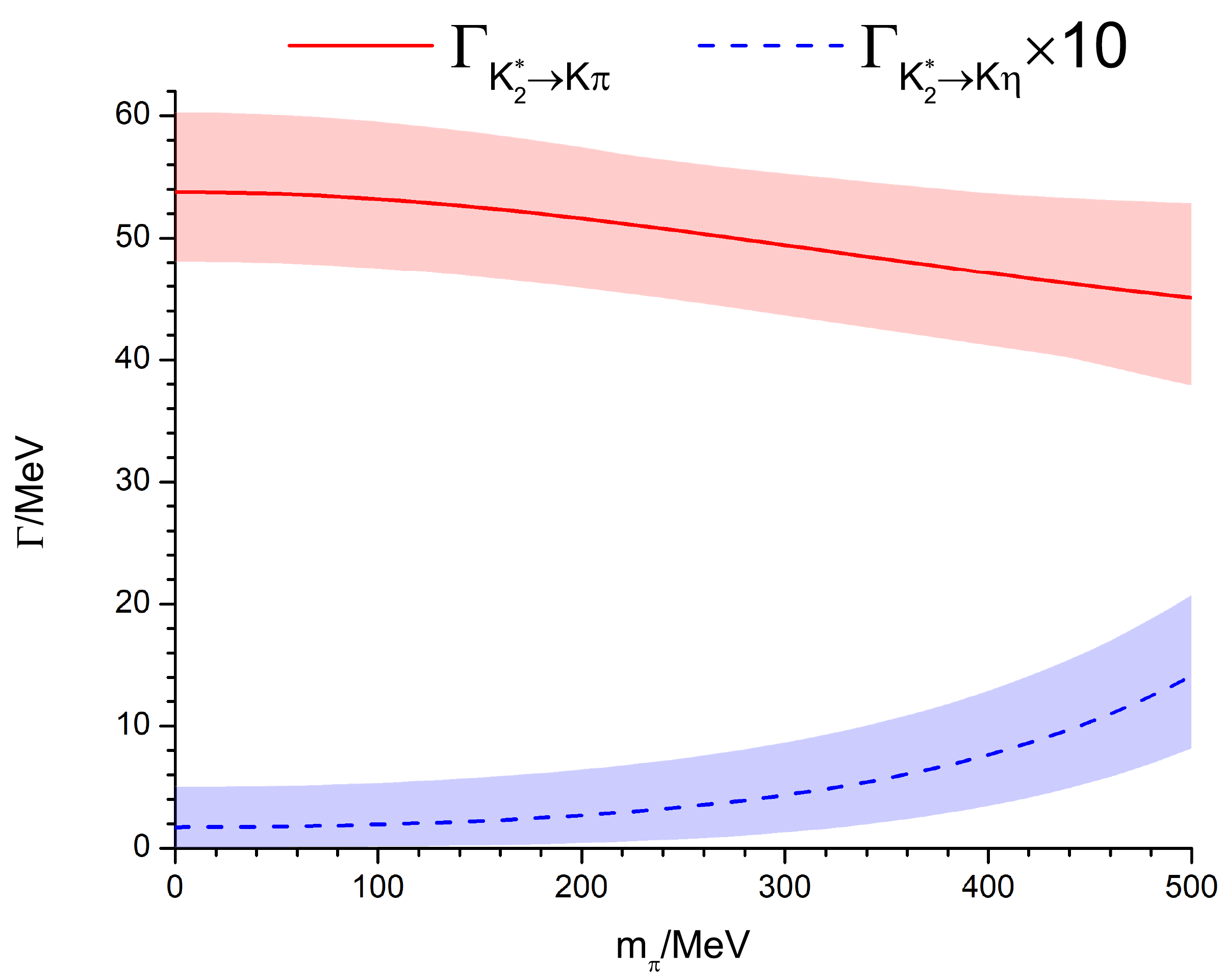} 
\caption{Partial two-meson decay widths of $a_2$ (left panel) and $K_2^*$ (right panel) as functions of $m_\pi$. The mass of strange quark is set at its physical value and the shaded areas correspond to our estimation of the theoretical uncertainties, see the text for details.}
 \label{fig.widtha2}
\end{figure}

In Figs.~\ref{fig.widthf2} and \ref{fig.widtha2}, we give predictions to the pion-mass dependence of the two-meson decay widths of the tensor resonances from the Fit-I scenario that are quoted in Table~\ref{tab.latwidth}. The shaded areas correspond to our estimation of theoretical uncertainties that are obtained by taking the parameters in Eq.~\eqref{eq.fiti}. As done in Fig.~\ref{fig.massandtheta}, we perform the extrapolation of $m_K$ via the LO $\chi$PT ansatz by fixing the mass of strange quark at its physical value. To give the predictions of the curves in Figs.~\ref{fig.widthf2} and \ref{fig.widtha2}, we also need to provide the extrapolations of the masses of $\eta$, $f_2$, $f_2'$, $a_2$ and $K_2^*$, and the mixing angles of $\eta$-$\eta'$ and $f_2$-$f_2'$, as functions of $m_\pi$. We use the NLO extrapolation result in Ref.~\cite{Gao:2022xqz} for the $m_\eta$. The pion-mass dependences of the tensor masses and the tensor mixing angle are taken from the discussions in Sec.~\ref{sec.massfit}, i.e. the curves in Fig.~\ref{fig.massandtheta}. For the extrapolation of the $\eta$-$\eta'$ mixing angle, we use the formula in Eq.~\eqref{eq.mixetalo}. In addition to the decay width of $K_2^*\to K \pi$, we also give the predictions in Fig.~\ref{fig.widtha2} to the pion-mass dependence of $\Gamma_{K_2^*\to K \eta}$, which turns out to be small in a broad range of $m_\pi$. The decay widths of the $f_2\to\pi\pi, K\bar{K}$, $f_2'\to\pi\pi, K\bar{K}$ and $a_2\to\pi\eta, K\bar{K}$ as functions of $m_\pi$ are explicitly shown in Figs.~\ref{fig.widthf2} and \ref{fig.widtha2}. These predictions of the two-meson decay widths at different values of $m_\pi$ can provide useful references for future lattice simulations.

\subsection{ $T\to P P'$ decay widths with two-mixing-angle formalism for $\eta$-$\eta'$ }

In this part, we apply the two-mixing-angle formula in Eq.~\eqref{eq.mixeta08} to study of the $T\to P P^\prime$ decay widths. Since the two-mixing-angle description results from taking into account the $\chi$PT contributions at least from NLO~\cite{Guo:2015xva} to the $\eta$-$\eta'$ system, the higher-order $\chi$PT effects should be also consistently introduced to $\pi$ and $K$, which requires us to distinguish their weak decay constants $F_\pi$ and $F_K$ in the $T\to P P'$ decay widths. As a result, for the two-mixing-angle scenario,  
we will set $F_\pi^{\rm Phy}=92.1$~MeV and $F_K^{\rm Phy}=110.0$~MeV~\cite{PDG2021} in the study of experimental widths. For the lattice decay widths, one should correspondingly extrapolate their values to the unphysically large meson masses. We will follow Ref.~\cite{Guo:2016zep} to set $F_\pi^{\rm Lat}=105.9$~MeV and $F_K^{\rm Lat}=115.0$~MeV when taking the lattice masses given in Eq.~\eqref{eq.massmpilat}.

For the two-mixing-angle scenario, there are four $\eta$-$\eta'$ mixing parameters~\eqref{eq.mixeta08}, i.e. $F_0, F_8, \theta_0$ and $\theta_8$. Due to the limited number of decay channels involving $\eta$ and $\eta'$ for the tensor resonances, our strategy is to greatly exploit the results from the recent NLO $U(3)$ study in Ref.~\cite{Gao:2022xqz} that incorporates a large amount of lattice data of the $\eta$-$\eta'$ mixing~\cite{Christ:2010dd,Gregory:2011sg,Dudek:2011tt,Ottnad:2017bjt,Bali:2021qem}. After exploration of different setups, our preferred fit is obtained by freeing the $\theta_8$ and fixing the values of $F_0, F_8$ and $\theta_0$ from Ref.~\cite{Gao:2022xqz} when analyzing the experimental $T\to P P^\prime$ decay widths. For the partial decay widths from lattice study, there is only one available decay channel involving $\eta$ or $\eta'$. It turns out that sensible fit can be reached by fixing all the four mixing parameters predicted in Ref.~\cite{Gao:2022xqz}. To be more specific, the values of the mixing parameters taken from the previous reference are $F_0^{\rm Phy}=97.0$~MeV, $F_8^{\rm Phy}=115.8$~MeV, $\theta_0^{\rm Phy}=-5.1^\circ$ at the physical masses, and $F_0^{\rm Lat}=102.3$MeV, $F_8^{\rm Lat}=117.4$MeV, $\theta_0^{\rm Lat}=-4.5^\circ$, $\theta_8^{\rm Lat}=-15.4^\circ$ at the lattice masses~\eqref{eq.massmpilat}. 

The resulting parameters from our preferred fit in the two-mixing-angle scenario are 
\begin{eqnarray}
&g_T= (19.9\pm 1.5)~{\rm MeV}\,, \quad &f_T= (1.2\pm 0.2)\times 10^{-5}~{\rm MeV^{-1}}\,, \nonumber \\
&g'_T= (6.3\pm 0.7)~{\rm MeV}\,, \quad &f'_T= (7.5\pm 5.3)\times 10^{-6}~{\rm MeV^{-1}}\,,\quad \theta^{\rm Phy}_8= (-17.3\pm 6.3)^\circ \,,  \nonumber \\
\end{eqnarray} 
with $\chi^2/({\rm d.o.f})= 13.4/(18-5)$. The value of resulting $\chi^2$ is similar to that of the one-mixing-angle fit discussed previously. In fact, we have also tried to use a common LO decay constant $F$ to replace $F_\pi$ and $F_K$ in the decay amplitudes involving pion and kaon, and in the meantime to impose the two-mixing-angle formalism for $\eta$ and $\eta'$. Nevertheless we find that such kind of setup leads to rather large $\chi^2$ in the fit and fails to simultaneously describe the $T\to P P'$ decay widths shown in Tables~\ref{tab.physwidth} and \ref{tab.latwidth}. Therefore our study here shows that it is crucial to consistently include the chiral corrections to the $\pi$, $K$, $\eta$, and $\eta'$ at the same order in order to obtain sensible results in the $T\to P P'$ decays.

\subsection{Radiative decays of tensor resonances}

The two main kinds of radiative decay processes for the tensor resonances are $T\to P \gamma$ and $T\to \gamma\gamma$.  One can similarly follow the R$\chi$T framework in Sec.~\ref{sec.theo} to construct the relevant operators to describe such processes. 

For the $T\to P\gamma$ decay, the relevant LO R$\chi$T operator can be written as~\cite{Kubis:2015sga} 
\begin{align}\label{eq.lagtpg}
 \mathscr{L}^{TP\gamma}=&  i\frac{c_{TP\gamma}}{2}\epsilon_{\mu\nu\alpha\beta}\langle T^{\alpha\lambda}[f_+^{\mu\nu},\partial^\beta u_\lambda]\rangle\,,
\end{align}
where $\epsilon_{\mu\nu\alpha\beta}$ is the antisymmetric tensor, and the chiral building blocks are  
\begin{align}
&f_{+}^{\mu \nu}=uF^{\mu\nu}u^{\dagger}+u^{\dagger}F^{\mu\nu}u\,,\qquad F^{\mu\nu}=-e\mathcal{Q}(\partial^\mu A^\nu-\partial^\nu A^\mu)\,,
\end{align}
with the quark electric-charge matrix: $\mathcal{Q}={\rm Diag} \left\{ \frac{2}{3},-\frac{1}{3},-\frac{1}{3}\right\}$. 
The $T\to P\gamma$  decay width from the LO Lagrangian~\eqref{eq.lagtpg} is 
\begin{align}
\Gamma_{T\rightarrow P\gamma}
=\frac{4\alpha c_{TP\gamma}^2}{5 F^2}\left(\frac{m_T^2-m_P^2}{2m_T} \right)^5\,,
\end{align}
where $\alpha=\frac{e^2}{4\pi}$ denotes the fine structure constant, $m_T$ and $m_P$ are the masses of the tensor and pseudoscalar mesons in the process of $T\to P\gamma$, respectively. $F$ corresponds to the weak decay constant of the pNGBs at chiral limit, which will be fixed at $F=81.7$~MeV as in the previous discussions. By taking the following experimental decay widths~\cite{PDG2021} as inputs
\begin{eqnarray}\label{eq.tpgwidthexp}
 \Gamma_{a_2^{\pm} \rightarrow\pi^{\pm}\gamma}^{\rm Exp}= (0.31\pm0.04)~{\rm MeV}\,, \qquad 
 \Gamma_{K_2^{*\pm}\rightarrow K^{\pm}\gamma}^{\rm Exp}= (0.24\pm0.06)~{\rm MeV} \,,
\end{eqnarray}
the fit gives 
\begin{eqnarray}
c_{T P\gamma}=(5.4\pm 0.5)\times10^{-5}~{\rm MeV^{-1}}\,.
\end{eqnarray}
This in turn leads to our theoretical results 
\begin{eqnarray}
  \Gamma_{a_2^{\pm}\rightarrow\pi^{\pm}\gamma}^{\rm Theo}= (0.30\pm0.04)~{\rm MeV}\,, \qquad 
 \Gamma_{K_2^{*\pm}\rightarrow K^{\pm}\gamma}^{\rm Theo}= (0.25\pm0.03)~{\rm MeV} \,,
\end{eqnarray}
which perfectly agree with the experimental inputs in Eq.~\eqref{eq.tpgwidthexp}. Therefore it is clear that the single R$\chi$T operator in Eq.~\eqref{eq.lagtpg} can already give a satisfactory description to the two available experimental decay widths of $a_2^{\pm}\to \pi^{\pm}\gamma$ and $K_2^{*\pm}\to K^{\pm}\gamma$.

For the $T\to \gamma\gamma$ processes, the available data from experiments~\cite{PDG2021} are 
\begin{eqnarray}\label{eq.tggwidthexp}
\Gamma_{f_2\to\gamma\gamma}^{\rm Exp}= 2.7\pm0.5 \,, \qquad 
\Gamma_{f_2'\to\gamma\gamma}^{\rm Exp}= 0.082\pm 0.015\,,\qquad  
\Gamma_{a_2\to\gamma\gamma}^{\rm Exp}= 1.0\pm0.1 \,, 
\end{eqnarray}
which are given in units of KeV. Following Ref.~\cite{Bellucci:1994eb}, the LO R$\chi$T operator that describes the $T\to\gamma\gamma$ process is given by 
\begin{eqnarray}\label{eq.lagtggc}
\mathscr{L}_{T\gamma\gamma}^{(0)}= c_{T\gamma\gamma}\langle T_{\mu\nu}\Theta_\gamma^{\mu\nu}\rangle\,,
\end{eqnarray}
with 
\begin{align}\label{eq.thetamunu}
\Theta_\gamma^{\mu\nu}=f_{+\alpha}^{\mu}f_{+}^{\alpha\nu}+\frac{1}{4}g^{\mu\nu}f_{+}^{\rho \sigma}f_{+\rho \sigma}\,. 
\end{align}
Since the on-shell tensor resonance is traceless~\cite{Ecker:2007us}, the second term in Eq.~\eqref{eq.thetamunu} will not contribute to the $T\to\gamma\gamma$ decay widths. At LO, the value of $c_{T\gamma\gamma}$ can be completely fixed by the  $a_2\to\gamma\gamma$ decay width, which leads to $c_{T\gamma\gamma}=2.4 \times 10^{-4}$~MeV$^{-1}$. By further taking this value of $c_{T\gamma\gamma}$ and the tensor mixing angle $\theta_T=29.0^\circ$ in Table~\ref{tab.mass}, we can give the LO prediction: $\Gamma_{f_2\to\gamma\gamma}^{\rm LO}=2.6$~KeV and $\Gamma_{f_2'\to\gamma\gamma}^{\rm LO}=0.12$~KeV, the latter of which is in clear tension with the experimental value in Eq.~\eqref{eq.tggwidthexp}. Next we try two possible ways to improve the simultaneous description for all the three types of $T\to\gamma\gamma$ decay widths. In one of them, we keep the value of $c_{T\gamma\gamma}=2.4 \times 10^{-4}$~MeV$^{-1}$ from the determination of $a_2\to\gamma\gamma$ and fit the tensor mixing angle $\theta_T$. This leads to $\theta_T=(27.2\pm1.1)^\circ$, which is in tension with the mixing angle $\theta_T=(29.0\pm 0.4)^\circ$ determined from the analyses of the tensor masses in Sec.~\ref{sec.massfit}. The other possible way is to introduce a new R$\chi$T operator by including the quark mass correction to the LO one:    
\begin{eqnarray}\label{eq.lagtggd}
\mathscr{L}_{T\gamma\gamma}^{(1)}= d_{T\gamma\gamma}\langle T_{\mu\nu}\Theta_\gamma^{\mu\nu}\chi_+\rangle\,.
\end{eqnarray}
The decay widths of various $T\to\gamma\gamma$ processes from the operators in Eqs.~\eqref{eq.lagtggc} and \eqref{eq.lagtggd} take the form 
\begin{align}
\Gamma_{a_2\to \gamma\gamma}=\left[\frac{4}{3\sqrt{2}}c_{T\gamma\gamma}
+\frac{4\sqrt{2}m^2_\pi}{3}d_{T\gamma\gamma}\right]^2 \frac{\pi \alpha^2 m_{a_2}^3}{20}\,,
\end{align}
\begin{align}
\Gamma_{f_2\to \gamma\gamma}=&\left[\frac{2\sqrt{2}\cos\theta_T+\sin\theta_T}{3\sqrt{6}}c_{T\gamma\gamma}
+\frac{4(m^2_K+2m^2_\pi)\cos\theta_T+\sqrt{2}(7m^2_\pi -4m^2_K)\sin\theta_T}{9\sqrt{3}}d_{T\gamma\gamma}
\right]^2\notag\\&\frac{4\pi \alpha^2 m_{f_2}^3}{5}\,,
\end{align}
\begin{align}
\Gamma_{f_2^\prime\to \gamma\gamma}=&\left[\frac{\cos\theta_T-2\sqrt{2}\sin\theta_T}{3\sqrt{6}}c_{T\gamma\gamma}
+\frac{\sqrt{2}(7m^2_\pi -4m^2_K)\cos\theta_T-4(m^2_K+2m^2_\pi)\sin\theta_T}{9\sqrt{3}}d_{T\gamma\gamma}
\right]^2\notag\\&\frac{4\pi \alpha^2 m_{f_2^\prime}^3}{5}\,.
\end{align}

Then we proceed the discussions by fixing the tensor mixing angle $\theta_T=29.0^\circ$ from the mass analysis and freeing the two couplings $c_{T\gamma\gamma}$ and $d_{T\gamma\gamma}$ to simultaneously fit the three decay widths in Eq.~\eqref{eq.tggwidthexp}. The fit gives 
\begin{align}
c_{T\gamma\gamma}=(2.4\pm 0.1)\times 10^{-4}~{\rm MeV^{-1}}\,, \qquad
d_{T\gamma\gamma}=(-3.2\pm 1.5)\times 10^{-11}~{\rm MeV^{-3}}\,,
\end{align}
which in turn leads to our theoretical results 
\begin{eqnarray} 
\Gamma_{f_2\to\gamma\gamma}^{\rm Theo}= 2.6\pm 0.2 \,, \qquad 
\Gamma_{f_2'\to\gamma\gamma}^{\rm Theo}= 0.082\pm 0.015\,,\qquad  
\Gamma_{a_2\to\gamma\gamma}^{\rm Theo}= 1.0\pm 0.1 \,,
\end{eqnarray}
which are given in units of KeV. The above values are now in perfect agreement with the experimental inputs in Eq.~\eqref{eq.tggwidthexp}, and in the meantime the tensor mixing angle appearing in the $T\to\gamma\gamma$ decay widths is completely fixed by the tensor mass analysis.

\section{Summary and conclusions}\label{sec.sum}

In this work we rely on resonance chiral theory to study the lowest tensor nonet, including $f_2, a_2, K_2^*$ and $f_2'$. Their masses and the decay widths from the $T\to PP'$, $T\to P\gamma$, and $T\to \gamma\gamma$ channels are calculated by including pertinent higher-order resonance chiral operators. By properly performing the chiral extrapolations at different quark masses, we have successfully fitted not only the relevant experimental data but also the corresponding lattice simulation results at unphysically large quark masses. 

Via the joint fit to the tensor masses from experiment and lattice, the $f_2$-$f_2'$ mixing angle is precisely determined to be $\theta_T=(29.0 \pm 0.4)^\circ$ at physical meson masses, which can be easily extrapolated to other unphysical masses. Apart from the leading-order interaction operator, we also construct higher-order Lagrangians by  including the quark-mass and $1/N_C$ corrections to study the various $T\to P P'$ decay processes from experiments and lattice simulations. In the fits to the $T\to P P'$ decay widths, we have fixed the tensor mixing angle $\theta_T$ from the mass analysis and exploited both one-mixing-angle and two-mixing-angle scenarios for the $\eta$-$\eta'$ mixing in the processes involving $\eta$ or $\eta'$. The qualities of the two fits turn out to be similar. 
Two different types of tensor radiative decays, i.e. $T\to P\gamma$ and $T\to\gamma\gamma$, are studied as well. For the $T\to P\gamma$ case, the two decay widths of $a_2^{\pm}\to \pi^{\pm}\gamma$ and $K_2^{*\pm}\to K^{\pm}\gamma$ can be well described by the single leading-order resonance chiral operator. While for the $T\to\gamma\gamma$ case, it is possible to simultaneously reproduce the three decay widths of $f_2\to\gamma\gamma$, $f_2'\to\gamma\gamma$ and $a_2\to\gamma\gamma$ with the single leading operator $c_{T\gamma\gamma}$ and tensor mixing angle $\theta_T$. However the resulting value of tensor mixing angle $\theta_T$ is somewhat inconsistent with the determination from the tensor mass analysis. Therefore our preferred description of the $T\to\gamma\gamma$ decays needs to introduce at least one extra higher-order resonance chiral operator in addition to the leading one, while fixing the tensor mixing angle from the mass analysis. Our studies in this work provide useful ingredients for future calculations of the tensor contributions to various scattering processes and also offer analytical tools to perform chiral extrapolations involving tensor resonances for future lattice simulations at different quark masses.

\section*{Acknowledgements}

We would like to thank Prof.~Liang Tang for useful discussions. This work is partially funded by the Natural Science Foundation of China (NSFC) under Grants No.~11975090, No.~12150013, and the Science Foundation of Hebei Normal University with Contract No.~L2023B09. 

\appendix

\section{The decay widths of $T\rightarrow P_1 P_2$ processes}
\setcounter{equation}{0}
\def\theequation{\Alph{section}.\arabic{equation}}\label{appendix.FFpsip}

By using the R$\chi$T Lagrangians in Eqs.~\eqref{eq.lagint0} and \eqref{eq.lagint1}, it is straightforward to calculate the $T\to P_1 P_2$ decay widths. Their explicit expressions are  
\begin{align}
\Gamma_{f_2\to\pi\pi}=& \left[\frac{4\cos\theta_T+2\sqrt{2}\sin\theta_T}{F_\pi ^2}g_T+
\frac{\left(16\cos\theta_T+8\sqrt{2}\sin\theta_T\right)m^2_\pi}{F_\pi^2}f_T
+\frac{6\cos\theta_T}{F_\pi^2}g_T^\prime\right.\notag\\
&+\left.
\frac{\left(8\cos\theta_T+4\sqrt{2}\sin\theta_T\right)m^2_\pi}{F_\pi^2}f_T^\prime\right]^2 \frac{p^5(m_{f_2},m_\pi,m_\pi)}{30\pi m_{f_2}^2}\,,
\end{align}
\begin{align}
\Gamma_{f_2\to KK}=&\left[\frac{8\cos\theta_T-2\sqrt{2}\sin\theta_T}{\sqrt{3}F_K^2}g_T+
\frac{32m^2_K\cos\theta_T-8\sqrt{2}\left(4m^2_K-3m^2_\pi\right)\sin\theta_T}
{\sqrt{3}F_K^2}f_T\right.\notag\\
&+\frac{12\cos\theta_T}{\sqrt{3} F_K^2}g_T^\prime+\left.\frac{16m^2_K\cos\theta_T+4\sqrt{2}\left(2m^2_K-3m^2_\pi\right)\sin\theta_T}
{\sqrt{3}F_K^2}f_T'\right]^2\frac{p^5(m_{f_2},m_K,m_K)}{30\pi m_{f_2}^2}\,,
\end{align}
\begin{align}
	&\Gamma_{f_2 \rightarrow \eta \eta}= \nonumber\\
	& \Bigg\{\frac { 2 } { 3 \sqrt { 3 }F _ { 0 } ^ { 2 } F _ { 8 } ^ { 2 } \cos^2 ( \theta _ { 0 } - \theta _ { 8 } )  } \bigg\{ -\sin \theta_T\left[\sqrt{2} F_0^2[3 g_T+2(2 f_T+f_T^{\prime})(8 m_K^2-5 m_\pi^2)]\cos ^2 \theta_0\right. \nonumber\\ 
	& +\left. 2 F_0 F_8[6 g_T+8 f_T(4 m_K^2-m_\pi^2)] \cos \theta_0 \sin \theta_8+8 \sqrt{2} F_8^2 (2 f_T+f_T^{\prime})(m_K^2-m_\pi^2) \sin ^2 \theta_8\right]\nonumber\\ 
	&+\cos \theta_T\Big[F _ { 0 } ^ { 2 } [6 g_T+8 f_T(4 m_K^2-m_\pi^2)] \cos ^2 \theta_0+16 \sqrt{2} F_0 F_8(2 f_T+f_T^{\prime})(m_K^2-m_\pi^2) \cos \theta_0 \sin \theta_8  \nonumber\\ 
	& +F_8^2(6 g_T+9 g_T^{\prime }+16 f_T m_K^2+8 f_T^{\prime} m_K^2+ 8 f_T m_\pi^2+4 f_T^{\prime} m_\pi^2) \sin ^2 \theta_8\Big]\bigg\}\Bigg\}^2 \frac{p^5(m_{f_2},m_\eta,m_\eta)}{30 \pi m_{f_2}^2}\,,
\end{align}
\begin{align}
\Gamma_{f_2^\prime\to \pi\pi}=&\left[\frac{-4\sin\theta_T+2\sqrt{2}\cos\theta_T}{F_\pi^2}g_T+
\frac{\left(-16\sin\theta_T+8\sqrt{2}\cos\theta_T\right)m^2_\pi}{F_\pi^2}f_T
-\frac{6\sin\theta_T}{F_\pi^2}g_T^\prime
\right.\notag\\
&\left.+
\frac{\left(-8\sin\theta_T+4\sqrt{2}\cos\theta_T\right)m^2_\pi}{F_\pi^2}f_T^\prime\right]^2\frac{p^5(m_{f_2^\prime},m_\pi,m_\pi)}{30\pi m_{f_2^\prime}^2}\,,
\end{align}
\begin{align}
\Gamma_{{f_2^\prime}\to KK}=&\left[\frac{8\sin\theta_T+2\sqrt{2}\cos\theta_T}{\sqrt{3}F_K^2}g_T+
\frac{8\sqrt{2}\left(4m^2_K-3m^2_\pi\right)\cos\theta_T+32m^2_K\sin\theta_T}
{\sqrt{3}F_K^2}f_T\right.\notag\\
&+\frac{12\sin\theta_T}{\sqrt{3}F_K^2}g_T^\prime\left.+\frac{16m^2_K\sin\theta_T-4\sqrt{2}\left(2m^2_K-3m^2_\pi\right)\cos\theta_T}
{\sqrt{3}F_K^2}f_T^\prime\right]^2\frac{p^5(m_{f_2^\prime},m_K,m_K)}{30\pi m_{f_2^\prime}^2}\,,
\end{align}
\begin{align}
	&\Gamma_{f_2^{\prime} \rightarrow \eta \eta}= \nonumber\\
	& \Bigg\{\frac { 2 } { 3 \sqrt { 3 } F _ { 0 } ^ { 2 } F _ { 8 } ^ { 2 } \cos^2 ( \theta _ { 0 } - \theta _ { 8 } )  } \bigg\{\cos \theta _ { T } \left[\sqrt{2} F_0^2[3 g_T+2(2 f_T+f_T^{\prime})(8 m_K^2-5 m_\pi^2)] \cos ^2 \theta_0\right. \nonumber\\ 
	& +\left. 2 F_0 F_8[6 g_T+8 f_T(4 m_K^2-m_\pi^2)] \cos \theta_0 \sin \theta_8+8 \sqrt{2} F_8^2(2 f_T+f_T^{\prime})(m_K^2-m_\pi^2) \sin ^2 \theta_8\right] \nonumber\\ 
	& +\sin \theta_T\left[F_0^2[6 g_T+8 f_T(4 m_K^2-m_\pi^2)] \cos ^2 \theta_0+16 \sqrt{2} F_0 F_8(2 f_T+f_T^{\prime})(m_K^2-m_\pi^2) \cos \theta_0 \sin \theta_8 \right.\nonumber\\ 
	& + F_8^2(6 g_T+9 g_T^{\prime }+16 f_T m_K^2 
	+8 f_T^{\prime} m_K^2+8 f_T m_\pi^2+4 f_T^{\prime} m_\pi^2) \sin ^2 \theta_8\Big]\bigg\}\Bigg\}^2 \frac{p^5(m_{f_2^\prime},m_\eta,m_\eta)}{30 \pi m_{f_2^{\prime }}^2} \,,
\end{align}
\begin{align}
\Gamma_{a_2\to KK}=\left[\frac{2\sqrt{2}}{F_K^2}g_T+\frac{8\sqrt{2}m^2_\pi}{F_K^2}f_T+
\frac{2\sqrt{2}\left(4m^2_K-2m^2_\pi\right)}{F_K^2}f_T^\prime\right]^2\frac{p^5(m_{a_2},m_K,m_K)}{30\pi m_{a_2}^2}\,,
\end{align}
\begin{align}
	\Gamma_{a_2 \rightarrow \pi \eta}= & \left\{\frac { 2 } { \sqrt { 3 }F_\pi F _ { 0 } F _ { 8 } \cos  ( \theta _ { 0 } - \theta _ { 8 } ) } \left\{\sqrt{2} F_0\left[g_T+2\left(2 f_T+f_T^{\prime}\right) m_\pi^2\right] \cos \theta_0\right.\right. \nonumber\\ 
	&- F_8\left[2 g_T+4\left(2 f_T+f_T^{\prime}\right) m_\pi^2\right] \sin \theta_8\Big\}\bigg\}^2 \frac{p^5(m_{a_2},m_\pi,m_{\eta})}{15 \pi m_{a_2}^2} \,,
\end{align}
\begin{align}
	\Gamma_{a_2 \rightarrow \pi \eta^\prime}= & \left\{\frac { 2 } { \sqrt { 3 }F_\pi F _ { 0 } F _ { 8 } \cos  ( \theta _ { 0 } - \theta _ { 8 } ) } \left\{\sqrt{2} F_0\left[g_T+2\left(2 f_T+f_T^{\prime}\right) m_\pi^2\right] \sin \theta_0\right.\right. \nonumber\\ 
	&+ F_8\left[2 g_T+4\left(2 f_T+f_T^{\prime}\right) m_\pi^2\right] \cos \theta_8\Big\}\bigg\}^2 \frac{p^5(m_{a_2},m_\pi,m_{\eta^\prime})}{15 \pi m_{a_2}^2} \,,
\end{align}
\begin{align}
\Gamma_{K_2^*\to K\pi}=\left[\frac{2\sqrt{3}}{F_\pi  F_K}g_T+\frac{8\sqrt{3}m^2_K}{F_\pi  F_K}f_T+\frac{4\sqrt{3}m^2_\pi}{F_\pi  F_K}f_T^\prime\right]^2\frac{p^5(m_{K_2^*},m_K,m_\pi)}{30\pi m_{K_2^*}^2}\,,
\end{align}
\begin{align}
	\Gamma_{K_2^* \rightarrow K \eta}= & \left\{\frac { 1 } { \sqrt { 3 }  F_K F _ { 0 } F _ { 8 } \cos  ( \theta _ { 0 } - \theta _ { 8 } ) } \left\{\sqrt{2} F_0\left[g_T+4 f_T m_K^2+f_T^{\prime}\left(8 m_K^2-6 m_\pi^2\right)\right] \cos \theta_0\right.\right. \nonumber\\ 
	&+2 F_8\left[2 g_T+4\left(2 f_T+f_T^{\prime}\right) m_K^2\right] \sin \theta_8\Big\} \bigg\}^2 \frac{p^5(m_{K_2^*},m_K,m_\eta)}{15 \pi m_{K_2^*}^2}\,,
\end{align}
where for the $T\to P_1 P_2$ process the three-momentum $p$ in the tensor rest frame is given by 
\begin{equation}
p(m_T,m_1,m_2)=\frac{\sqrt{[m_T^2-(m_1+m_2)^2][m_T^2-(m_1-m_2)^2]}}{2m_T}\,.
\end{equation}
It is noted that we have introduced different $F_\pi$ and $F_K$ factors for  different decay processes involving pion and kaon. In the one-mixing-angle scenario, both $F_\pi$ and $F_K$ reduce to the chiral limit value $F$.

\end{document}